\documentstyle[seceq,epsf,preprint]{jpsj}
\def\runtitle{Quasiparticle States near the Surface and the Domain Wall 
in a $p_x$$\pm$i$p_y$-Wave Superconductor}     
\def\runauthor{Masashige {\sc Matsumoto} and Manfred {\sc Sigrist}}

\title
{Quasiparticle States near the Surface and the Domain Wall 
in a $p_x$$\pm$i$p_y$-Wave Superconductor}

\author
{Masashige {\sc Matsumoto} and Manfred {\sc Sigrist}$^1$}

\inst
{Department of Physics, Faculty of Science, Shizuoka University, 836 Oya,
Shizuoka 422-8529, Japan \\
$^1$Yukawa Institute for Theoretical Physics, Kyoto University, Kyoto 606-8502, Japan}

\recdate
{}
\abst
{The electronic states near a surface or a domain wall
in the $p$-wave superconductor are studied for the order parameter of
the form $p_x$$\pm$i$p_y$-wave, which is a unitary odd-parity state with
broken time-reversal symmetry. 
This state has been recently suggested as the superconducting state of
Sr$_2$RuO$_4$. 
The spatial variation of the order parameter and vector potential is
determined self-consistently within the quasi-classical approximation.
The local density of states at the surface is constant and does not
show any peak-like or gap-like structure within the superconducting
energy gap, in contrast to the case of the $d$-wave
superconductors. The influence of an external magnetic field is mainly 
observable in the energy range above the bulk gap. 
On the other hand, there is a small energy gap in the local density of
states at the domain wall between domains of the two degenerate
$p_x$+i$p_y$-wave  and  $p_x$$-$i$p_y$-wave states.}

\kword
{Sr$_2$RuO$_4$, $p$-wave superconductor, unitary state, time-reversal breaking state,
boundary effect, surface, domain wall, quasi-classical theory, Ginzburg-Landau free energy}

\begin{document}
\sloppy
\maketitle
\newcommand{\no}{\noindent}
\newcommand{\beq}{\begin{equation}}
\newcommand{\beqn}{\begin{eqnarray}}
\newcommand{\eeq}{\end{equation}}
\newcommand{\eeqn}{\end{eqnarray}}
\newcommand{\ri}{{\rm i}}
\newcommand{\fx}{{\rm F}_x}
\newcommand{\vfx}{v_{{\rm F}x}}
\newcommand{\vfy}{v_{{\rm F}y}}
\newcommand{\vfxa}{v_{{\rm F}x1}}
\newcommand{\vfxb}{v_{{\rm F}x2}}
\newcommand{\delx}{{\partial \over \partial x}}
\newcommand{\dely}{{\partial \over \partial y}}
\newcommand{\br}{\mbox{\boldmath$r$}}
\newcommand{\rd}{{\rm d}}
\newcommand{\bk}{\mbox{\boldmath$k$}}
\newcommand{\kh}{{\hat k}}
\newcommand{\tk}{{\theta_k}}
\newcommand{\bhz}{\hat {\mbox{\boldmath$z$}}}
\newcommand{\kf}{k_{\rm F}}
\newcommand{\kfx}{{k_{{\rm F}x}}}
\newcommand{\kfy}{{k_{{\rm F}y}}}
\newcommand{\kfxa}{{k_{{\rm F}x1}}}
\newcommand{\kfxb}{{k_{{\rm F}x2}}}
\newcommand{\bkf}{{\mbox{\boldmath$k$}}_{\rm F}}
\newcommand{\om}{{\omega_{m}}}
\newcommand{\Ps}{{\hat \Psi}}
\newcommand{\Ph}{{\hat \Phi}}
\newcommand{\Pha}{\Ph_\alpha}
\newcommand{\Phb}{\Ph_\beta}
\newcommand{\Del}{{\hat \Delta}}
\newcommand{\Dela}{\Delta_\alpha}
\newcommand{\Delb}{\Delta_\beta}
\newcommand{\Delah}{{{\hat \Delta}_\alpha}}
\newcommand{\Delbh}{{{\hat \Delta}_\beta}}
\newcommand{\tauc}{{\hat \tau_3}}
\newcommand{\taub}{{\hat \tau_2}}
\newcommand{\G}{{\hat G}}
\newcommand{\hg}{{\hat g}}
\newcommand{\ab}{{\alpha \beta}}
\newcommand{\U}{{\hat U}}
\newcommand{\TU}{{\tilde U}}
\newcommand{\Ua}{{{\hat U}_\alpha}}
\newcommand{\Ub}{{{\hat U}_\beta}}
\newcommand{\A}{{\hat A}}
\newcommand{\B}{{\hat B}}
\newcommand{\Lam}{{\hat \Lambda}}
\newcommand{\re}{{\rm e}}
\newcommand{\bd}{\mbox{\boldmath{$d$}}}
\newcommand{\bR}{\mbox{\boldmath{$R$}}}
\newcommand{\bX}{\mbox{\boldmath{$X$}}}
\newcommand{\hDelta}{\hat \Delta}
\newcommand{\Deltauu}{\hat \Delta_{\uparrow\uparrow}}
\newcommand{\Deltaud}{\hat \Delta_{\uparrow\downarrow}}
\newcommand{\Deltadu}{\hat \Delta_{\downarrow\uparrow}}
\newcommand{\Deltadd}{\hat \Delta_{\downarrow\downarrow}}
\newcommand{\hpsi}{\hat \psi}
\newcommand{\hPhi}{\hat \Phi}
\newcommand{\hpsiu}{\hat \psi_{\uparrow}}
\newcommand{\hpsid}{\hat \psi_{\downarrow}}
\newcommand{\hG}{\hat G}
\newcommand{\hF}{\hat F}
\newcommand{\bq}{\mbox{\boldmath{$q$}}}
\newcommand{\hsigma}{\hat \sigma}
\newcommand{\htau}{\hat \tau}
\newcommand{\bhsigma}{\hat {\mbox{\boldmath{$\sigma$}}}}
\newcommand{\bn}{\boldmath{\mbox{$n$}}}
\newcommand{\bp}{\mbox{\boldmath{$p$}}}
\newcommand{\Q}{\mbox{\boldmath{$Q$}}}
\newcommand{\bx}{\mbox{\boldmath{$x$}}}
\newcommand{\hkf}{{\hat k}_{\rm F}}
\newcommand{\vf}{v_{\rm F}}
\newcommand{\bvf}{\mbox{\boldmath{$v_{\rm F}$}}}
\newcommand{\bA}{\mbox{\boldmath{$A$}}}
\newcommand{\psiu}{\psi_\uparrow}
\newcommand{\psid}{\psi_\downarrow}
\newcommand{\ep}{\varepsilon}
\renewcommand{\dag}{\dagger}
\newcommand{\taua}{\hat {\tau_1}}
\newcommand{\sigb}{\hat {\sigma_2}}
\newcommand{\del}{\partial}
\newcommand{\ome}{\omega_{\rm m}}
\newcommand{\bfkp}{\mbox{\boldmath $k_+$}}
\newcommand{\bfkn}{\mbox{\bolkmath $k_-$}}
\newcommand{\bskf}{\mbox{\footnotesize \boldmath $k_{\rm F}$}}
\newcommand{\bfsq}{\mbox{\footnotesize \boldmath $q$}}
\newcommand{\bfskp}{\bfsk_+}
\newcommand{\bfskn}{\bfsk_-}
\newcommand{\bfsqp}{\bfsq_+}
\newcommand{\bfsqn}{\bfsq_-}
\newcommand{\rhok}{\rho_\bfsk}
\newcommand{\rhokp}{\rho_\bfsk}
\newcommand{\xperp}{x_\perp}
\newcommand{\ypara}{y_\parallel}
\newcommand{\ky}{k_y}
\newcommand{\m}{\hat \mu}
\newcommand{\n}{\hat \nu}
\newcommand{\phai}{\hat \varphi}
\newcommand{\pha}{\hat \phi}
\newcommand{\qy}{q_y}
\newcommand{\sig}{\hat \Sigma}
\newcommand{\tsigb}{\Sigma_2}
\newcommand{\tsigc}{\Sigma_3}
\newcommand{\fbar}{\overline{f}}
\newcommand{\dbar}{D_x}
\newcommand{\Ef}{{E_{\rm F}}}
\newcommand{\ubar}{\bar u}
\newcommand{\vbar}{\bar v}
\newcommand{\rpara}{r_\parallel}
\newcommand{\fig}[1]
{\vspace{24pt}
\begin{center}
\fbox{\rule{0cm}{#1}\hspace{7cm}}
\end{center}}

\section{Introduction}
Sr$_2$RuO$_4$ is the first superconductor with layered perovskite
structure, which does not contain copper. \cite{Maeno} Although
the structure is identical to that of some of the high-temperature
superconductors, the transition temperature is rather low, $T_{\rm C}$=1.5K.
There is a clear difference in the electronic structure, since
Sr$_2$RuO$_4$ is a good metal and even a Fermi liquid in its
stoichiometric composition. Band structure calculations in good
agreement with the de Haas-van Alphen measurements show that this compound
has three Fermi surfaces originating from the three $ 4d $-$t_{2g}
$-orbitals of Ru$^{4+}$. \cite{Mackenzie1,Oguchi,Singh} There is
growing experimental evidence that the 
superconducting state is unconventional (non-s-wave). Examples are the absence
of a Hebel-Slichter peak in $ 1/T_1 T $ of NQR-measurements
\cite{Ishida}
and the sensitivity of $ T_{\rm C} $ on non-magnetic impurities \cite{Mackenzie2}. 

It was suggested that the superconducting state has odd-parity
(spin triplet) pairing. \cite{Rice,Bask,Sigrist1,Machida,Mazin} There is a
certain similarity with $^3$He considering
the correlation effects (superfluid $^3$He has $p$-wave
pairing). \cite{Rice} Furthermore, there is a series of related
compounds such as  
SrRuO$_3$ which are ferromagnetic suggesting that
ferromagnetic spin fluctuations are probably enhanced in 
Sr$_2$RuO$_4$ and mediate odd-parity, spin triplet
pairing. \cite{Gibb,Rice,Sigrist6} The recent
discovery of intrinsic magnetism in the superconducting phase by $\mu
$SR experiments indicates a pairing
state with broken time reversal symmetry. \cite{Luke} Symmetry
considerations lead to the conclusion that this would only be possible 
for an odd-parity state. \cite{Sigrist5} 
A very strong support for odd-parity pairing
comes also from the Knight shift data in the $^{17}$O-NMR measurements 
which demonstrate the absence of any reduction of the spin
susceptibility in the superconducting state. \cite{Ishida2} The
superconducting state  
compatible with all of these experiments is given by $\bd(\bk)$=$\bhz(k_x$$\pm$i$k_y)$.
\cite{Sigrist5}  

The presence of three electron bands forming the Fermi liquid state
leads to the question whether all of them contribute to superconductivity 
on equal terms. Symmetry considerations show that it is possible to
separate the orbitals $ \{ 4 d_{yz} , 4 d_{zx} \} $ (forming two Fermi 
surfaces) and $ 4 d_{xy} $ (forming one Fermi surface). 
Thus it is theoretically possible that superconductivity predominantly 
appears in one of the two sets of orbitals, while the other
participates passively only through induced superconductivity. \cite{Agterberg1}
There are several supporting experiments for the scenario of
``orbital dependent superconductivity''.
The specific heat
\cite{Nishizaki}
and NQR experiments
\cite{Ishida}
indicate a considerable residual density of states in the 
superconducting state which could be attributed to the passive
electron bands. Furthermore, the analysis of the London penetration
depth extrapolated to the zero temperature is compatible with
taking only one of the two subsets of orbitals into account for the
superconducting state. \cite{Riseman} The present experimental situation is
compatible with a superconducting state dominantly in the $4d_{xy}$-orbital.
The recent prediction of a square form of the vortex
lattice aligned with the crystal lattice axis was confirmed
experimentally and is consistently explained with the assumption of
superconductivity in the $ 4 d_{xy} $-orbital as
well. \cite{Agterberg2,Riseman} It seems 
therefore to be justified to approximate the electronic structure by a 
single band model consisting a single practically cylinder-shaped
Fermi surface.

The theoretical study of boundary effects in high-temperature
superconductors with $d$-wave Cooper pairing have turned out to be very
fruitful for subsequent experimental
investigations. 
The bound states cause unusual features in the Andreev reflection and
tunneling spectroscopy. 
\cite{Hu,Kashiwaya,Ohashi,Matsumoto1,Nagato1,Buchholtz,Higashitani2}
Recently characteristic tunneling conductance properties were examined 
theoretically for the triplet pairing states. 
\cite{Honerkamp,Yamashiro1} Conductance peak features related to
the bound states are very sensitive to the angle of the incidence in
this case, in contrast to the $d$-wave superconductor.
In this paper we investigate the electronic states near the surface and
compare them with those found near domain walls. Domain walls appear due 
to the fact that the superconducting state is two-fold degenerate 
($p_x$+i$p_y$- and $p_x$$-$i$p_y$-wave state) such 
that two types of domains are possible. Because of the discrete degeneracy
the separating domain walls are very well localized and
constitute a region of spatial deformation of the order parameter
giving rise to subgap quasiparticle states analogous to the surface.

\section{The Quasiclassical Formulation}
In this paper we study boundary effects in the $p$-wave superconductor,
determining the order parameters self-consistently.
For this purpose we use the quasi-classical Green function formalism
developed by Schopohl, which, for example, 
has been applied to the vortex problem. \cite{Schopohl1,Ichioka}
We adapt this scheme to the boundary problem and apply it to both the
case of the surface and the domain wall.

For simplicity we assume that the superconductor is two-dimensional
with a single cylindrical Fermi surface which is a reasonable first
approximation to the band of the $4d_{xy}$-orbital,
and that the superconducting order parameter has a $p$-wave symmetry.
The gap matrix for the $p$-wave state is given by the $\bd$-vector
defined as
\beq
\hDelta(\br,\br')=\ri\bd(\br,\br')\cdot\bhsigma\hsigma_y,
\eeq
where $\hsigma_i$$(i$=$x,y,z)$ is the Pauli spin matrix.
We restrict ourselves to the case of
$\bd(\br,\br')$=$(0,0,\Delta(\br,\br'))$ type, so that the 
Bogoliubov-de Gennes equation reduces to a $2\times 2$ matrix form,
\beq
{\displaystyle \int}\rd\br'
\left(
  \matrix{
    \delta(\br-\br')h_0(\br') & \Delta(\br,\br') \cr
    -\Delta^*(\br,\br') & -\delta(\br-\br')h_0^*(\br') \cr
  }
\right)
\left(
  \matrix{
    u_l(\br') \cr
    v_l(\br') \cr
  }
\right)
=
E_l
\left(
  \matrix{
    u_l(\br) \cr
    v_l(\br) \cr
  }
\right).
\eeq
Here $\Delta(\br,\br')$ is the $p$-wave pair potential,
$E_l$ is the $l$-th energy eigenvalue, and 
$h_0(\br)$=[$-$$\ri\nabla$+$e\bA(\br)]^2/2m$$-$$\Ef$ is the kinetic energy of an electron
measured from the Fermi energy, with $e$ ($e$$>$0) and $\bA$ as the
charge of the electron and vector potential, respectively. 
For convenience $\hbar$ is taken to be unity throughout this paper.

Introducing the Andreev approximation (i.e. separating the rapid oscillations) as
\beq
\left(
  \matrix{
    u_l(\br) \cr
    v_l(\br) \cr
  }
\right)=
\left(
  \matrix{
    \ubar_l(\bkf,\br) \cr
    \vbar_l(\bkf,\br) \cr
  }
\right)
\re^{\ri\bskf\cdot\br},
\eeq
we obtain the Andreev equation
\cite{Andreev,Bruder},
\beq
\left(
  \matrix{
    -\ri\bvf\cdot\nabla+e\bvf\cdot\bA(\br) & \Delta(\bkf,\br) \cr
    -\Delta^*(-\bkf,\br) & \ri\bvf\cdot\nabla+e\bvf\cdot\bA(\br) \cr
  }
\right)
\left(
  \matrix{
    \ubar_l(\bkf,\br) \cr
    \vbar_l(\bkf,\br) \cr
  }
\right)
=
E_l
\left(
  \matrix{
    \ubar_l(\bkf,\br) \cr
    \vbar_l(\bkf,\br) \cr
  }
\right),
\label{eqn:2.4}
\eeq
where $\bkf$ and $\bvf$ are the Fermi wave number and the Fermi velocity,
respectively.
The $\bkf$ dependence of $\Delta(\bkf,\br)$ represents the symmetry of
the order parameter 
and $\br$ is the center position of the Cooper pair.
Due to the $p$-wave symmetry (i.e. $\Delta($$-$$\bkf,\br)$=$-$$\Delta(\bkf,\br)$),
eq. (\ref{eqn:2.4}) has the same form as the Andreev equation in the case
of singlet pairing. The corresponding Eilenberger equation is given by
\beq
-\ri\bvf\cdot\nabla\hg(\bkf,\om,\br)=
\left[
  \left(
    \matrix{
      \ri\om-e\bvf\cdot\bA(\br) & -\Delta(\bkf,\br) \cr
      \Delta^*(\bkf,\br) & -\ri\om+e\bvf\cdot\bA(\br) \cr
    }
  \right),
  \hg(\bkf,\om,\br)
\right],
\eeq
where $\hg(\bkf,\om,\br)$ is the quasi-classical Green function in a
$2\times 2$ matrix form and
$\om$=$\pi T(2m$+1) ($m$: integer) is the fermion Matsubara
frequency.\cite{Eilenberger,Serene,Ashida} 
For simplicity the Boltzmann constant is taken to be unity.
The equations for the matrix elements of the quasi-classical Green
function are expressed as 
\beqn
&&\hg(\bkf,\om,\br)=
\left(
  \matrix{
    g(\bkf,\om,\br) & \ri f(\bkf,\om,\br) \cr
    -\ri\fbar(\bkf,\om,\br) & -g(\bkf,\om,\br) \cr
  }
\right), \cr
&&\bvf\cdot\nabla g(\bkf,\om,\br)=
\Delta^*(\bkf,\br)f(\bkf,\om,\br)-\Delta(\bkf,\br)\fbar(\bkf,\om,\br), \cr
&&(\om+\ri e\bvf\cdot\bA(\br)+{1 \over 2}\bvf\cdot\nabla)f(\bkf,\om,\br)=
\Delta(\bkf,\br)g(\bkf,\om,\br), \cr
&&(\om+\ri e\bvf\cdot\bA(\br)-{1 \over 2}\bvf\cdot\nabla)\fbar(\bkf,\om,\br)=
\Delta^*(\bkf,\br)g(\bkf,\om,\br).
\label{eqn:2.6}
\eeqn
These equations contain the following symmetries,
\beqn
g(-\bkf,\om,\br)&=&g^*(\bkf,\om,\br),~~~~~~~~~
g(\bkf,-\om,\br)=-g^*(\bkf,\om,\br), \cr
f(-\bkf,\om,\br)&=&-\fbar^*(\bkf,\om,\br),~~~~~~
f(\bkf,-\om,\br)=\fbar^*(\bkf,\om,\br).
\label{eqn:2.7}
\eeqn
Using the following transformation,
we can solve the Eilenberger equation in a simple way,
\cite{Schopohl1}
\beqn
g(\bkf,\om,\br)&=&
{1-a(\bkf,\om,\br)b(\bkf,\om,\br) \over 1+a(\bkf,\om,\br)b(\bkf,\om,\br)}, \cr
f(\bkf,\om,\br)&=&{2a(\bkf,\om,\br) \over 1+a(\bkf,\om,\br)b(\bkf,\om,\br)}, \cr
\fbar(\bkf,\om,\br)&=&{2b(\bkf,\om,\br) \over 1+a(\bkf,\om,\br)b(\bkf,\om,\br)},
\label{eqn:2.8}
\eeqn
where $a$ and $b$ satisfy
\beqn
\bvf\cdot\nabla a(\bkf,\om,\br)&=&
\Delta(\bkf,\br)-\Delta^*(\bkf,\br)a^2(\bkf,\om,\br)
-2\bigl[\om+\ri e\bvf\cdot\bA(\br)\bigr]a(\bkf,\om,\br), \cr
\bvf\cdot\nabla b(\bkf,\om,\br)&=&
-\Delta^*(\bkf,\br)+\Delta(\bkf,\br)b^2(\bkf,\om,\br)
+2\bigl[\om+\ri e\bvf\cdot\bA(\br)\bigr]b(\bkf,\om,\br).
\nonumber \\
\label{eqn:2.9}
\eeqn

Let us now consider the case of a surface perpendicular to the
$x$-direction. We assume that quasiparticles are specularly reflected
at the surface. 
We solve eq. (\ref{eqn:2.9}) along the classical trajectories as shown
in Fig. \ref{fig:1}(a) where the quasiparticle moves with the momentum 
close to $\bk_{\rm F1}$ from A to B and with momentum $\bk_{\rm F2}$
from B to C. Since the surface is specular and translational
invariant along the $y $- and $z $-direction,
the incident momentum along the surface is conserved.
We then match the two solutions using
the boundary condition of the quasi-classical Green function at point
B given by 
\cite{Serene,Ashida}
\beq
\hg({\bkf}_1,\om,{\rm B})=\hg({\bkf}_2,\om,{\rm B}),
\eeq
which for $a$ and $b$ means,
\beq
a({\bkf}_1,\om,{\rm B})=a({\bkf}_2,\om,{\rm B}),~~~~~~
b({\bkf}_1,\om,{\rm B})=b({\bkf}_2,\om,{\rm B}).
\label{eqn:2.11}
\eeq

Since the system is translationally invariant perpendicular to the $ x 
$-axis, the quasi-classical Green function
or $a(\bkf,\om,\br)$ and $b(\bkf,\om,\br)$ depend only on $x$.
Thus eq. (\ref{eqn:2.9}) can be rewritten as
\beqn
{\rd \over \rd x} a(\bkf,\om,x)&=&
{1 \over \vfx}
\Bigl\{
  \Delta(\bkf,x)-\Delta^*(\bkf,x)a^2(\bkf,\om,x)
  -2\bigl[\om+\ri e\vfy A_y(x)\bigr] a(\bkf,\om,x)
\Bigr\}, \cr
{\rd \over \rd x}b(\bkf,\om,x)&=&
{1 \over \vfx}
\Bigl\{
  -\Delta^*(\bkf,x)+\Delta(\bkf,x)b^2(\bkf,\om,x)
  +2\bigl[\om+\ri e\vfy A_y(x)\bigr] b(\bkf,\om,x)
\Bigr\}.
\nonumber \\
\label{eqn:2.12}
\eeqn
where $\vfx$ and $\vfy$ are the $x$ and $y$ component of the Fermi velocity,
respectively.
The initial and boundary conditions for eq. (\ref{eqn:2.12}) are given by
\beqn
a({\bkf}_1,\om,x=\infty)&=&
{\Delta({\bkf}_1,\infty) \over \sqrt{[\om+\ri e\vfy A_y(\infty)]^2
+\vert\Delta({\bkf}_1,\infty)\vert^2}+\om+\ri e\vfy A_y(\infty)}, \cr
b({\bkf}_2,\om,x=\infty)&=&
{\Delta^*({\bkf}_2,\infty) \over \sqrt{[\om+\ri e\vfy A_y(\infty)]^2
+\vert\Delta({\bkf}_2,\infty)\vert^2}+\om+\ri e\vfy A_y(\infty)}, \cr
a({\bkf}_1,\om,x=0)&=&a({\bkf}_2,\om,x=0), \cr
b({\bkf}_1,\om,x=0)&=&b({\bkf}_2,\om,x=0).
\label{eqn:2.13}
\eeqn

The position-dependent order parameter can be determined
by the quasi-classical Green function.
The gap equations for the $p_x$- and $p_y$-wave
(i.e. $\Delta(\bkf,x)$=$\Delta_x(x)\cos\tk$+$\Delta_y(x)\sin\tk$)
are given (using eq. (\ref{eqn:2.7})) as
\cite{Bruder,Kieselmann}
\beqn
\left(
  \matrix{
    \Delta_x(x) \cr
    \Delta_y(x) \cr
  }
\right)&=&
\pi T{\displaystyle \sum_{\vert\om\vert<\omega_{\rm C}}}{1 \over 2\pi}
{\displaystyle \int_{-\pi}^\pi}\rd\tk
\left(
  \matrix{
    2V_p\cos\tk \cr
    2V_p\sin\tk \cr
  }
\right)
f(\tk,\om,x) \cr
&=&
\pi TV_p
{\displaystyle \sum_{0<\om<\omega_{\rm C}}}{1 \over \pi}{\displaystyle \int_0^\pi}\rd\tk
\left(
  \matrix{
    2\cos\tk \cr
    2\sin\tk \cr
  }
\right)
\bigl[f(\tk,\om,x)+\fbar^*(\tk,\om,x)\bigr] \cr
&=&
\pi TV_{\rm p}
{\displaystyle \sum_{0<\om<\omega_{\rm C}}}{1 \over \pi}
{\displaystyle \int_0^{\pi \over 2}}\rd\tk \cr
&\times&
\left(
  \matrix{
    2\cos\tk\bigl[
              f(\tk,\om,x)+\fbar^*(\tk,\om,x)
              -f(\pi-\tk,\om,x)-\fbar^*(\pi-\tk,\om,x)
            \bigr] \cr
    2\sin\tk\bigl[
              f(\tk,\om,x)+\fbar^*(\tk,\om,x)
            +f(\pi-\tk,\om,-x)+\fbar^*(\pi-\tk,\om,-x)
            \bigr] \cr
  }
\right), \cr
V_p&=&
{1 \over \log{T \over T_{\rm C}}+{\displaystyle \sum_{0<m<{\omega_{\rm C}/2\pi T}}}
{1 \over m-{1/2}}},
\label{eqn:2.14}
\eeqn
where $\tk$ is the angle of the Fermi momentum measured from the
$k_x$-axis and $\omega_{\rm C}$ is the cutoff energy. We write 
$\Delta_x$ and $\Delta_y$ for the order parameter components
corresponding to the $p_x$- and $p_y$-wave, respectively.
Generally, they are complex numbers.
$T_{\rm C}$ is the superconducting transition temperature
which we assume is the same for both components.

The supercurrent due to the broken time-reversal symmetry is a general
property, which was previously studied in the framework of
Ginzburg-Landau (GL) theory. \cite{Volovic,Sigrist4}
Here we investigate it within the quasi-classical approximation.
The current density along the $y$ direction
can be expressed by the quasi-classical Green function, 
\beqn
J_y(x)&=&-e\vf N(0)T{\displaystyle \sum_{\vert\om\vert<\omega_{\rm C}}}
{1 \over 2\pi}{\displaystyle \int_{-\pi}^\pi}\rd\tk\sin\tk(-\ri\pi)g(\tk,\om,x) \cr
&=&
-2e\vf N(0)T{\displaystyle \sum_{0<\om<\omega_{\rm C}}}
{\displaystyle \int_0^{\pi \over 2}}\rd\tk\sin\tk{\rm Im}
\bigl[g(\tk,\om,x)+g(\pi-\tk,\om,x)\bigr],
\nonumber \\
\label{eqn:2.15}
\eeqn
where the symmetry relations of eq. (\ref{eqn:2.7}) have been
used.\cite{Nagato,Higashitani} 
$N(0)$ is the normal state density of states per unit volume at $E_{\rm F}$ and
$n$ is the density of electrons, whereby both 
$N(0)$ and $n$ include up and down spin electrons.

The magnetic field and vector potential are calculated using the
Maxwell equation. 
The magnetic field and the vector potential can be determined as follows:
\beqn
B_z(x)&=&-\mu\int_0^x\rd x' J_y(x'),~~~~~~
A_y(x)=\mu\int_0^x\rd x' B_z(x')-\mu\int_0^\infty\rd x' B_z(x'),
\label{eqn:2.16}
\eeqn
with $\mu$ as the permeability.
Equation (\ref{eqn:2.16}) is chosen to satisfy the boundary conditions of
$B_z(0)$=0, $A_y(\infty)$=0.
Now we can solve the gap equation (\ref{eqn:2.14}) iteratively
by using eqs. (\ref{eqn:2.8}), (\ref{eqn:2.12}), (\ref{eqn:2.13}), (\ref{eqn:2.15})
and (\ref{eqn:2.16}) until the self-consistency is achieved.

\section{Quasiparticle Properties at the Surface}

\subsection{Self-consistent solution}
Before examining the boundary effect,
let us study first the bulk case, where
$f$ and $\fbar$ in the bulk are given by
\beqn
f(\tk,\om,{\rm bulk})&=&\fbar^*(\tk,\om,{\rm bulk})=
{\Delta(\tk,{\rm bulk}) \over \sqrt{\om^2+\vert\Delta(\tk,{\rm bulk})\vert^2}}, \cr
\Delta(\tk,{\rm bulk})&=&\Delta_x({\rm bulk})\cos\tk+\ri\Delta_y({\rm bulk})\sin\tk.
\label{eqn:3.1}
\eeqn
Here we have assumed real values for $\Delta_x$ and $\Delta_y$,
since we are interested in the $p_x$+i$p_y$-wave state.
Substituting eq. (\ref{eqn:3.1}) into the gap equation,
we can obtain the bulk solution at $T$=0:
\beq
\Delta_x({\rm bulk})=\Delta_y({\rm bulk})=\Delta(0)=2\omega_{\rm C}\re^{-{1
\over V_p}},~~~ 
T_{\rm C}={2\re^\gamma\omega_{\rm C} \over \pi}\re^{-{1 \over V_p}},~~~
{\Delta(0) \over T_{\rm C}}=\pi\re^{-\gamma}\simeq 1.76.
\eeq
Here $\gamma$ is the Euler's constant: $\gamma$=0.57721$\cdots$.
This solution is similar to the $s$-wave case,
since $\vert\Delta(\tk,{\rm bulk})\vert$ is a constant on the Fermi surface.
If one of the two order parameter components is absent,
the solution becomes
\beqn
\Delta_x({\rm bulk})&=&0,~~~
\Delta_y({\rm bulk})=\Delta_{\rm single}(0)=4\omega_{\rm C}\re^{-({1 \over
V_p}+{1 \over 2})}, \cr 
T_{\rm C}&=&{2\re^\gamma\omega_{\rm C} \over \pi}\re^{-{1 \over V_p}},~~~
{\Delta_{\rm single}(0) \over T_{\rm C}}=2\re^{-{1 \over
2}}\pi\re^{-\gamma}\simeq 2.14. 
\eeqn
Note that $\Delta_{\rm single}(0)$ is larger than $\Delta(0)$. 
This is due to the fact that there are gapless points on the Fermi surface
for the single existing solution and that the denomitators of $f$ and
$\fbar$ become small. 
Though $\Delta(0)$ is small,
its solution is more stable, since it opens a gap everywhere on the
Fermi surface. \cite{Sigrist6}

\begin{figure}[b]
\begin{center}
\begin{minipage}[t]{7cm}
\epsfxsize=7cm
\epsfbox{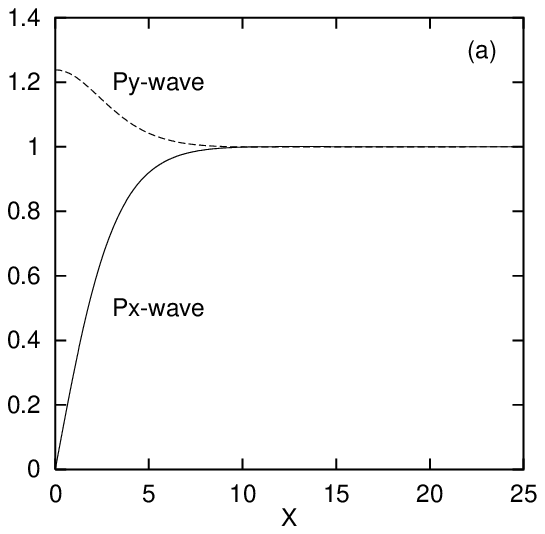}
\end{minipage}
\begin{minipage}[t]{7cm}
\epsfxsize=7cm
\epsfbox{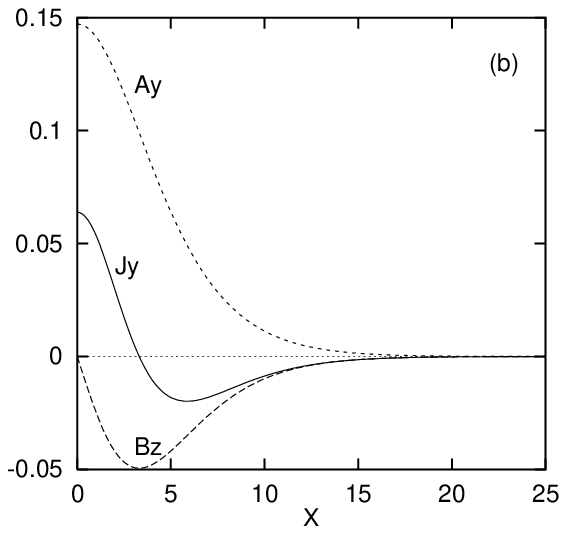}
\end{minipage}
\end{center}
\caption{
(a)~Spatial dependence of the self-consistent order parameters for
$p_x$- and $p_y$-waves. 
They are scaled by the bulk value $\Delta(T)$.
The $x$ coordinate is also scaled by $\xi_0$=${\vf /\pi\Delta(0)}$,
where $\Delta(0)$ is the magnitude of the bulk order parameter at $T$=0.
$\Delta_x$ and $\Delta_y$ are real and imaginary,
respectively.
The set of parameters are chosen as $T$=0.2$T_{\rm C}$, $\omega_{\rm C}$=10$T_{\rm C}$,
$\kappa$=$\lambda_{\rm L}/\xi_0$=2.5 (see ref. \ref{ref:2}).
Here $\lambda_{\rm L}$=$\sqrt{m/e^2\mu n}$ is the London penetration depth.
(b)~The self-consistent current density $J_y(x)$, magnetic field $B_z(x)$
and vector potential $A_y(x)$.
They are scaled by $J_0$=$e\vf N(0)T_{\rm C}$,
$B_{\rm C}$=$\Phi_0/2\sqrt{2}\pi\xi_0\lambda_{\rm L}$ and $\Delta(0)/e\vf$,
respectively.
Here $B_{\rm C}$ and $\Phi_0$=$h/2e$ are the critical magnetic field and flux quantum,
respectively.
}
\label{fig:2}
\end{figure}

We now turn to the surface problem.
As in the case of the $d$-wave pairing,
boundary effects can also be expected for $p$-wave pairing.
\cite{Honerkamp,Yamashiro1}
Due to the boundary condition eq. (\ref{eqn:2.11}) the gap equation
(\ref{eqn:2.14}) shows 
$\Delta_x$=0 at $x$=0. This is related to the fact that the sign of
the $p_x$-wave order parameter changes after the reflection at the surface.
In Fig. \ref{fig:2} we show the self-consistently obtained order parameter,
the current density, the magnetic field and the vector potential.
In the bulk region $\Delta_x$ and $\Delta_y$ take the same magnitude
as expected. 
Even if we start from an arbitrary relative phase, it 
finally becomes $\pm$${\pi/2}$ yielding the $ p_x$$\pm$i$p_y $-wave state.
In the GL theory this effect is included in the following fourth order term
of the GL 
free energy 
which is derived in the Appendix,
\begin{equation}
A_4
\Bigl[
  2\vert\Delta_x\vert^2\vert\Delta_y\vert^2
  +{1 \over 2}(\Delta_x^2{{\Delta_y}^*}^2+{{\Delta_x}^*}^2\Delta_y^2)
\Bigr].
\end{equation}
The first term represents the competition between $\Delta_x$ and $\Delta_y$,
while the second term favors $\pm$${\pi/2}$ for their relative phase.
Since the coefficient of the first term is four times larger than the
second one, 
$p_x$- and $p_y$-waves have to compete anyway.
Near the boundary  $\Delta_y$ is enhanced as shown in Fig. \ref{fig:2}(a).
This enhancement is connected with the suppression of $\Delta_x$ near
the surface, which opens the way for $ \Delta_y $ to appear as a ``single'' 
order parameter. However, we find that
$\Delta_y$ at the boundary (about 1.24) is larger
than $\Delta_{\rm single}(0)$
($\Delta_{\rm single}(0)/\Delta(0)$$\simeq$1.21).
Note that similar enhancement also appears in $d$-wave superconductors.
\cite{Bruder,Matsumoto1}
The reason can be seen in the following sixth order term of the GL
free energy, 
\beq
-4A_6 \vert\Delta_y\vert^2\vert\dbar\Delta_x\vert^2.
\eeq
This term supports the $ \Delta_y $-component whenever $ \Delta_x $
has a spatial variation and explains why $\Delta_y$$>$$\Delta_{\rm single}$ near the surface. 
Microscopically the enhancement of $\Delta_y$ near the surface
originates from the enhanced local density of states due to
additional (bound) states appearing at low energy close to the surface 
as a result of the depletion of $\Delta_x$ (as we will see below).
Obviously only the $ p_y $-component can benefit from this additional
density of states near the Fermi energy, since it does not suffer from
pair breaking for the surface perpendicular to the $ x$-axis.

Due to the broken time-reversal symmetry,
the local current flows along the surface as shown in
Fig. \ref{fig:2}(b), whose direction is reversed if we change the
relative phase of the order parameter components 
($p_x$+i$p_y$$\leftrightarrow$$p_x$$-$i$p_y$) (time reversal operation). 
The current along the $x$-direction is zero as expected.
The magnetic field is induced by the surface current
and it is screened by the Meissner effect,
so that the total current becomes zero.
\cite{Ohashi2} Note that depending on the geometry of the sample these 
currents can generate a finite magnetization.

\subsection{Density of states}
Due to the pair breaking effects
we expect a modification of the local density of states near the boundary.
The Andreev reflection and the conductance for the
triplet pairing states were studied previously by various groups that
the features of the conductance peak, which are related to the bound
states, depend strongly on the angle of the incidence.
\cite{Honerkamp,Yamashiro1}
The angle-resolved tunneling spectroscopy is related to the local
density of states 
for a fixed Fermi momentum which is obtained from the quasi-classical Green
function, 
\beq 
N(\tk,E,x)={\rm Re}\bigl[
                     g(\tk,\om,x)\mid_{\ri\om\rightarrow E+\ri\delta}
                   \bigr],
\label{eqn:3.6}
\eeq
where $\delta$ is a positive infinitesimal real number.
Before examining the self-consistent order parameters case,
it is illustrative to start with the consideration 
of a uniform order parameter, neglecting also the vector potential.
In this case we can obtain the quasi-classical Green function analytically,
\cite{Bruder,Matsumoto1}
\beqn
\hg(\tk,\om,x)&=&g_x(\tk,\om,x)\htau_x+g_y(\tk,\om,x)\htau_y+g_z(\tk,\om,x)\htau_z, \cr
g_x(\tk,\om,x)&=&g_x(\pi-\tk,\om,x)=
{-\Delta_y(\tk) \over \Omega}
\bigl[
  1+{\Delta_x^2(\tk) \over \om^2+\Delta_y^2(\tk)}\re^{-2q x}
\bigr]
-\ri{\om\vert\Delta_x(\tk)\vert \over \om^2+\Delta_y^2(\tk)}\re^{-2q x}, \cr
g_y(\tk,\om,x)&=&-g_y(\pi-\tk,\om,x)=
{-\Delta_x(\tk) \over \Omega}\bigl[
                               1-\re^{-2q x}
                             \bigr], \cr
g_z(\tk,\om,x)&=&g_z(\pi-\tk,\om,x)=
{\om \over \Omega}+
 {\om\Delta_x^2(\tk)-\ri\Omega\vert\Delta_x(\tk)\vert\Delta_y(\tk)
    \over \Omega[\om^2+\Delta_y^2(\tk)]}\re^{-{2q x}}, \cr
\Delta_x(\tk)&=&\Delta(T)\cos\tk,~~~
\Delta_y(\tk)=\Delta(T)\sin\tk,~~~
\Omega=\sqrt{\om^2+\Delta^2(T)},~~~
q={\Omega \over \vert\vfx\vert}.
\nonumber \\
\label{eqn:3.7}
\eeqn
Here $\htau_i$ ($i$=$x,y,z$) is the Pauli matrix in the charge space.
We have assumed $p_x$+i$p_y$-wave symmetry,
i.e.
$\Delta(\tk)$=$\Delta_x(\tk)$+i$\Delta_y(\tk)$=$\Delta(T)(\cos\tk$+i$\sin\tk$), 
where $\Delta(T)$ is the magnitude of the order parameter in the bulk
region at temperature $T$. 
At low temperatures we obtain $\Delta(T)$$\simeq$$\Delta(0)$.
Note that the enhancement of $\Delta_y$ near the boundary in
Fig. \ref{fig:2}(a) 
can be understood by the first term of $g_x$ 
\beq
\Delta_y(\tk)
\bigl[
  1+{\Delta_x^2(\tk) \over \om^2+\Delta_y^2(\tk)}\re^{-2q x}
\bigr]
>\Delta_y(\tk).
\eeq
The vanishing of $\Delta_x$ at the boundary can be derived from the equation for $g_y$.
Substituting $g_z$ into eq. (\ref{eqn:3.6}),
we have the following expression for the local density of states,
\beqn
N(\tk,E,x)&=&N(\pi-\tk,E,x) \cr
&=&
\bigl[
{\vert E\vert \over \rho}
-
{\Delta_x^2(\tk)\vert E\vert\cos(2\eta x)
+{\rm sgn}(E)\vert\Delta_x(\tk)\vert\Delta_y(\tk)\rho\sin(2\eta x)
 \over
\rho[E^2-\Delta_y^2(\tk)]}
\bigr]
\theta\bigl(E^2-\Delta^2(T)\bigr) \cr
&+&
\pi\vert\Delta_x(\tk)\vert
{\rm exp}\bigl({-{2\vert\Delta_x(\tk)\vert \over \vert\vfx\vert}x}\bigr)
\delta\bigl(E-\Delta_y(\tk)\bigr), \cr
\rho&=&\sqrt{E^2-\Delta^2(T)},~~~~~~\eta={\rho \over \vert\vfx\vert}.
\label{eqn:3.9}
\eeqn
The first term of eq. (\ref{eqn:3.9}) represents the continuum
quasiparticle states,
while the second term describes the bound state.
For $\Delta_y$=0 the bound state is located at $E$=0 while
a finite $p_y$-wave part leads to $E$=$\Delta_y(\tk)$=$\Delta(T)\sin\tk$.
\cite{Honerkamp}
This breaks the symmetry with respect to $k_y$$\rightarrow$$-$$k_y$,
so that the quasiparticle bound states with positive and negative $k_y$ have different occupation
at low temperatures.
Consequently, the quasiparticles generate a finite local surface current flows.
For the uniform order parameter
it is easy to see that $J_y(x)$ has a finite value at $x$=0
due to the $\vert\Delta_x\vert\Delta_y$ term of $g_z$ in
eq. (\ref{eqn:3.7}), 
at $T$=0 $J_y$(0)=$e\vf N(0){1 \over 2}\Delta(0)$.
This shows that the number of quasiparticles contributing to
the current is of the order of $N(0)\Delta(0)$.

\begin{figure}
\begin{center}
\begin{minipage}[t]{9cm}
\epsfxsize=9cm
\epsfbox{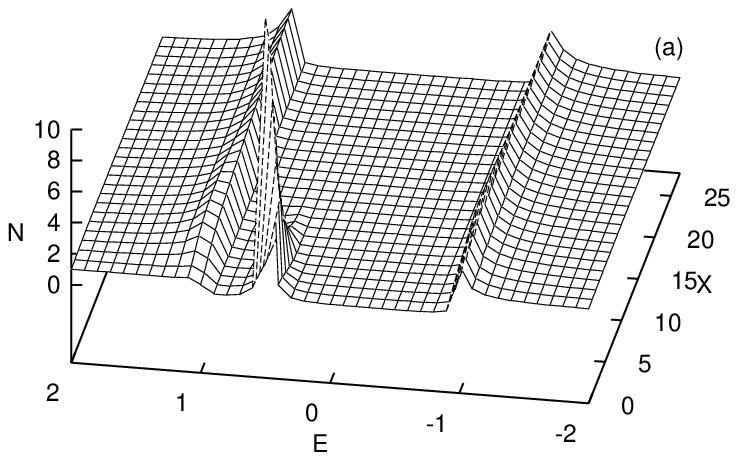}
\end{minipage}
\begin{minipage}[t]{7cm}
\epsfxsize=7cm
\epsfbox{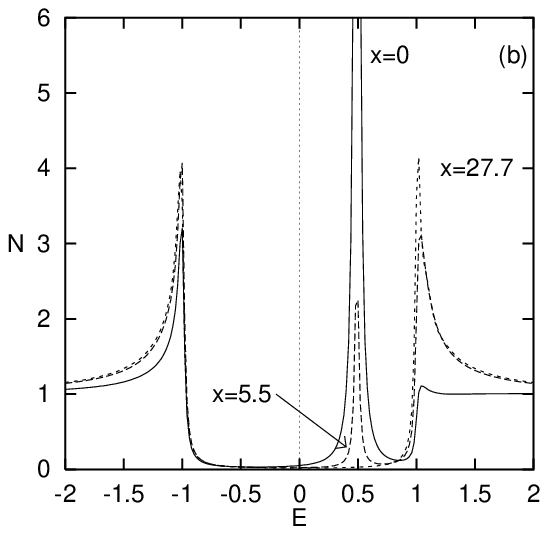}
\end{minipage}
\end{center}
\caption{
(a)~Local density of states for a fixed momentum,
which has been calculated from the self-consistent order parameters.
It is normalized by the one in the normal state.
The energy $E$ is scaled by $\Delta(T)$.
The momentum is chosen to give $\tk$=$\pi/8$.
A small imaginary part of 0.02$\times\Delta(T)$ is added to $E$
for the plotting.
(b)~Local density of states at various positions.
}
\label{fig:3}
\end{figure}

\begin{figure}
\begin{center}
\begin{minipage}[t]{9cm}
\epsfxsize=9cm
\epsfbox{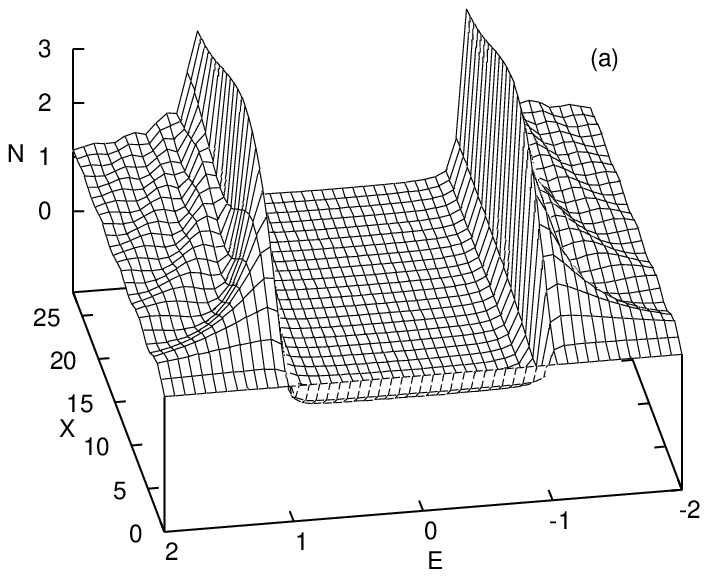}
\end{minipage}
\begin{minipage}[t]{7cm}
\epsfxsize=7cm
\epsfbox{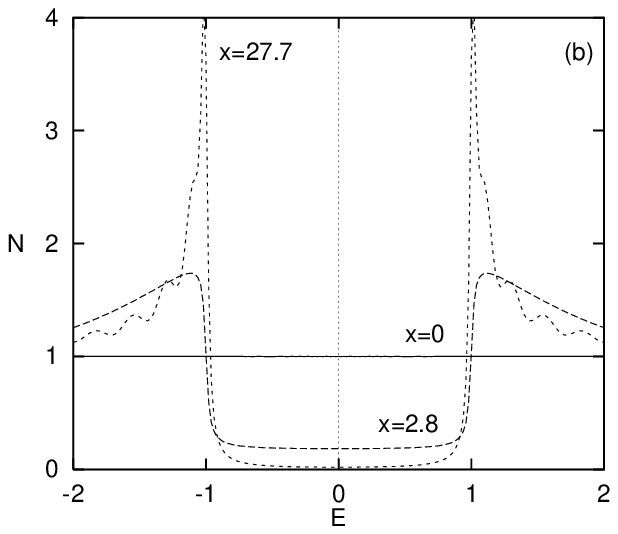}
\end{minipage}
\begin{minipage}[t]{9cm}
\epsfxsize=9cm
\epsfbox{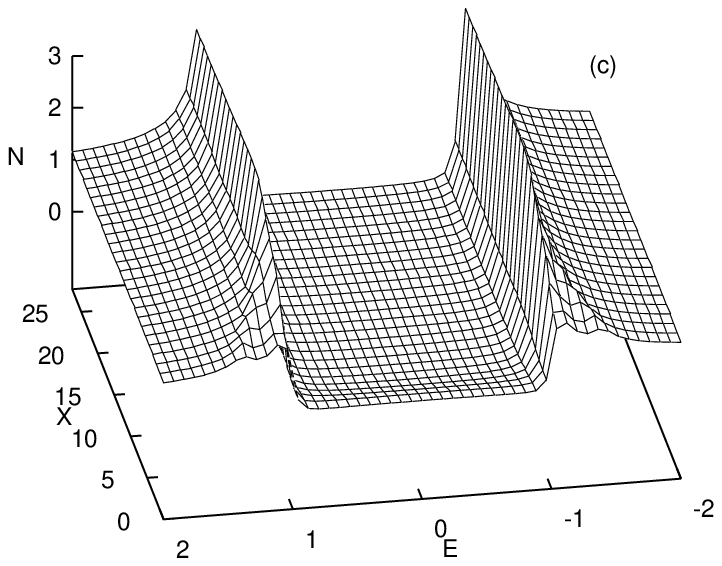}
\end{minipage}
\begin{minipage}[t]{7cm}
\epsfxsize=7cm
\epsfbox{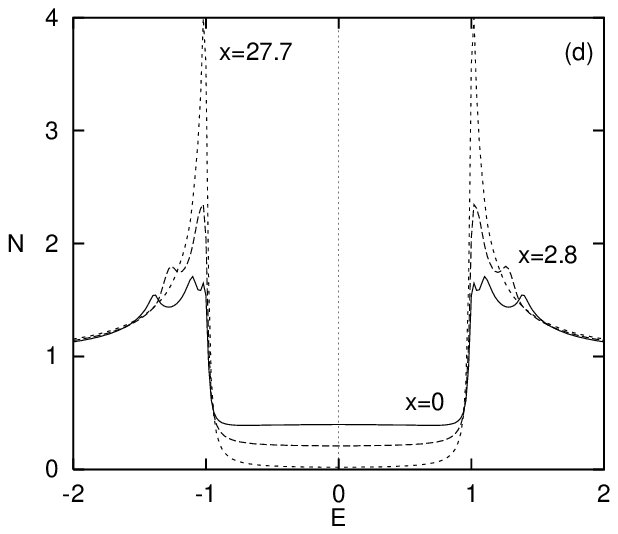}
\end{minipage}
\end{center}
\caption{
(a)~Total local density of states for uniform order parameters.
The vector potential is neglected for simplicity.
(b)~Total local density of states at various positions of (a).
(c)~Total local density of states for the self-consistent order parameters,
in which the vector potential is taken into account.
(d)~Total local density of states at various positions of (c).
}
\label{fig:4}
\end{figure}

We now turn to the results of the full selfconsistent calculation. 
Figure \ref{fig:3} shows the local density of states for a fixed
momentum $k_y$$>$0 ($\tk$=$\pi/8$) from the self-consistently determined order
parameter and vector potential. 
The energy of the bound state estimated from eq. (\ref{eqn:3.9}) is
sin$(\pi/8)\Delta(T)$$\simeq$0.38$\Delta(T)$.
In the $k_y$$<$0 case this energy is reversed in sign,
since $g$ has the symmetry of eq. (\ref{eqn:2.7}).
Therefore the integrated density of states over the Fermi surface
is symmetric under $E$$\rightarrow$$-$$E$.
In the case of the uniform order parameters the total local density of
states at the boundary 
exactly becomes the same as the normal state,
i.e. constant.
One can test this by integrating eq. (\ref{eqn:3.9}) over $\tk$ at $x$=0.
We show the local density of states calculated from eq. (\ref{eqn:3.9})
in Figs. \ref{fig:4}(a) and (b).
When we move away from the boundary,
the bound states rapidly decreases and the bulk property is recovered.

\begin{figure}
\begin{center}
\input{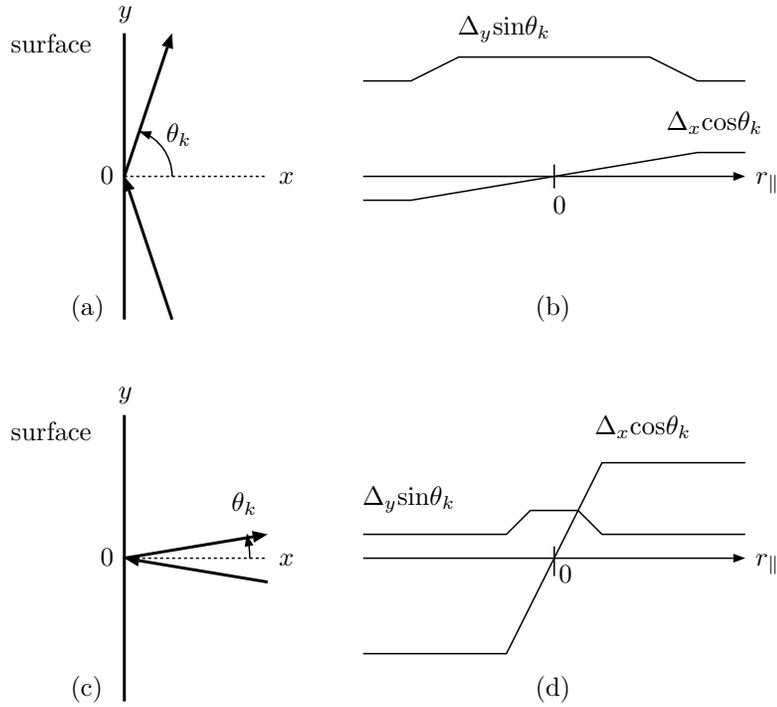}
\end{center}
\caption{
(a)~A classical trajectory of an quasiparticle reflected by a surface.
(b)~A schematic spatial dependence of the self-consistent order parameters
along the classical trajectory.
$\rpara$ is taken along the classical trajectory.
(c)~Same as (a) for a smaller $\tk$.
(d)~Same as (b) for a smaller $\tk$.
}
\label{fig:6b}
\end{figure}

On the other hand, we notice gap structures in Fig. \ref{fig:4}(d) even at $x$=0
for the self-consistent order parameter.
At the surface we can see two gap energies, 
one is $E/\Delta(T)$$\simeq$1, and the other forms a double peak
feature around about 1.25. 
The first one corresponds to the bulk gap energy,
while the second one is identified as the magnitude of $\Delta_y$ at $x$=0,
which is enhanced near the boundary.
These higher energy peaks are mainly formed by quasiparticles moving almost
parallel to the surface 
as shown in Fig. \ref{fig:6b}(a),
since they experience the enhanced $\Delta_y$ over a long distance
(see Fig. \ref{fig:6b}(b)).
According to eq. (\ref{eqn:2.4}) the energy should be replaced by
$E$$\rightarrow$$E$$-$$e\vfy A_y$ in the presence of a vector potential. This yields an
energy shift in the density of states. 
Quasiparticles running almost parallel to the surface
are strongly affected by the vector potential due to the factor $\vfy A_y$.
Without the vector potential, the higher energy peaks merge into
a single peak. Hence, the vector potential leads to the splitting into 
two peaks corresponding to the left and right moving particles along the surface.
On the other hand, 
the bound states are mainly formed by the quasiparticle moving almost
perpendicular to the surface 
shown in Figs. \ref{fig:6b}(c) and (d), which are only weakly affected 
by the vector potential.
The self-consistently determined $\Delta_x$ shows a continuous spatial
change near the surface which leads to a reduction of the magnitude of the
delta function peak corresponding to the 
bound state in $ N(\theta_k, E, x) $   
compared to the uniform order parameter case. The reason is the
widening of the potential well in which the bound state
quasiparticle is trapped. 
The spatial dependence of $\Delta_x$ can be approximately expressed by
tanh($x/(r\vf/\Delta(T))$), 
where $r$ controls the decay length of $\Delta_x$.
The magnitude of the bound state is then estimated by
\cite{Matsumoto1}
\beq
I(r)={\Gamma(r+1/2) \over \Gamma(1/2)\Gamma(1+r)}.
\eeq
Here $\Gamma$ is the gamma function, and
$I(r)$ takes 1 for $r$=0 and decreases with the increase of $r$.
For the self-consistent order parameter $r$ and $I(r)$ can be
estimated as 1 and 0.5, 
respectively.
Therefore the density of states of the bound state at $x$=0 in
Fig. \ref{fig:4}(d) 
is about half of Fig. \ref{fig:4}(b).

Consequently, the local density of states near the boundary is sensitive
to the spatial dependence of the order parameter,
so that the self-consistent treatment is needed for quantitative discussions.
However, the approximation of the uniform order parameters is helpful
to capture the essence of the boundary effect.

\section{External Magnetic Field}
In this section we study properties of the surface
in an applied external magnetic field $B_{\rm ext}$ to the $p_x$+i$p_y$-wave state.
In this case the boundary condition for the magnetic field is
$B_z(0)$=$B_{\rm ext}$, 
so that the first equation of (\ref{eqn:2.16}) has to be replaced by
\beq
B_z(x)=B_{\rm ext}-\mu\int_0^x\rd x' J_y(x').
\eeq

\begin{figure}
\begin{center}
\begin{minipage}{7cm}
\epsfxsize=7cm
\epsfbox{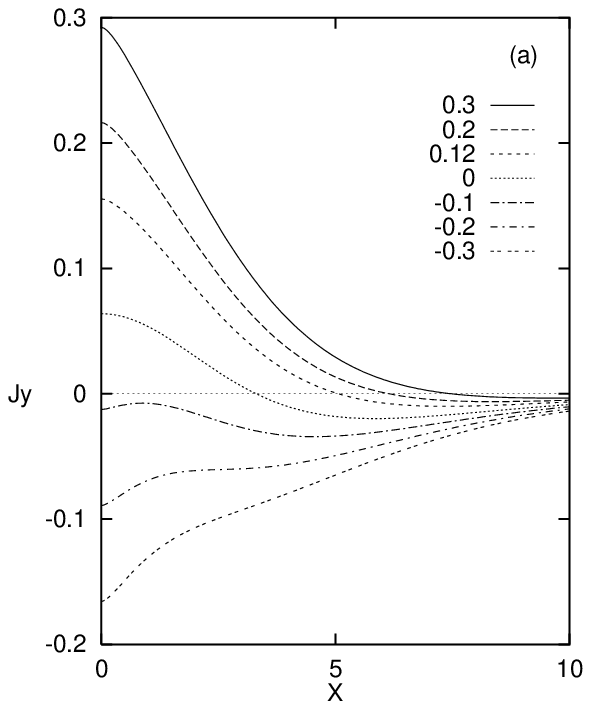}
\end{minipage}
\begin{minipage}{7cm}
\epsfxsize=7cm
\epsfbox{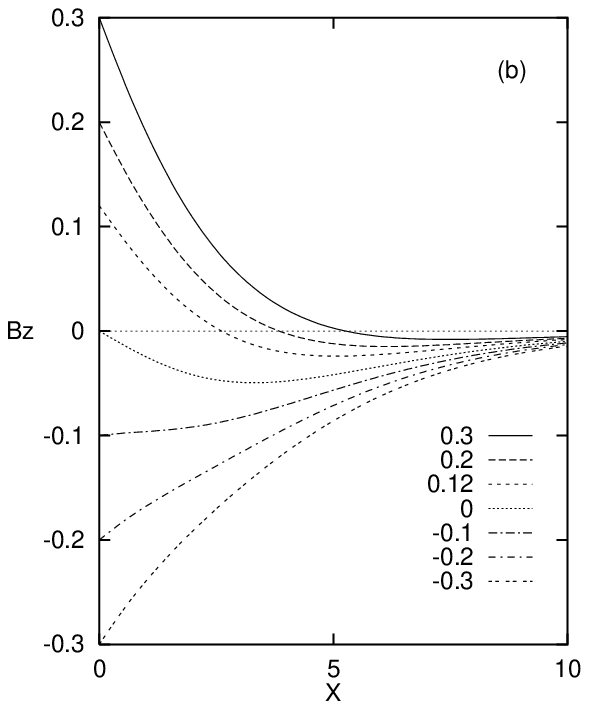}
\end{minipage}
\begin{minipage}{7cm}
\epsfxsize=7cm
\epsfbox{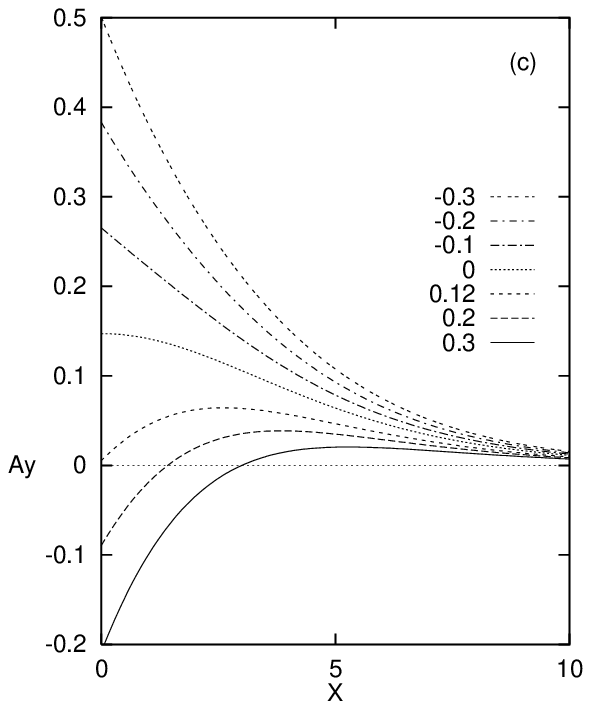}
\end{minipage}
\begin{minipage}{7cm}
\epsfxsize=7cm
\epsfbox{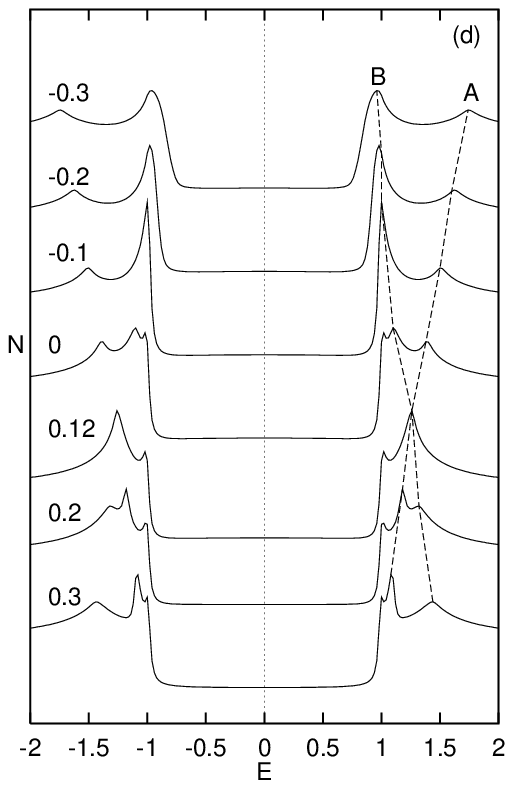}
\end{minipage}
\end{center}
\caption{
(a)~Spatial dependence of the self-consistent current $J_y(x)$
which flows along the surface.
$J_y(x)$ and $x$ are scaled by $J_0$=$e\vf N(0)T_{\rm C}$ and $\xi_0$,
respectively.
The parameters are chosen as $T$=0.2$T_{\rm C}$, $\omega_{\rm C}$=10$T_{\rm C}$,
$\kappa$=2.5.
(b)~Spatial dependence of the self-consistent magnetic field $B_z(x)$,
which is scaled by the critical field $B_{\rm C}$=$\Phi_0/2\sqrt{2}\pi\xi_0\lambda_{\rm L}$.
(c)~Spatial dependence of the self-consistent vector potential $A_y(x)$,
which is scaled by $\Delta(0)/e\vf$.
(d)~Local density of states at $x$=0 in an arbitrary unit.
In each figure the applied external magnetic fields normalized by $B_{\rm C}$ are depicted.
In the preset case $B_{c1}$ is estimated as
$B_{c1}/B_{\rm C}$=log$\kappa/\sqrt{2}\kappa$$\simeq$0.26.
}
\label{fig:8}
\end{figure}

We show the results of the self-consistent solution in Fig. \ref{fig:8}.
The order parameter does not suffer a significant modification
compared with the situation without the external field. 
In Fig. \ref{fig:8}(d) we can see that the peak position,
which corresponds to the enhanced $\Delta_y$,
shifts in a higher (lower) energy region if we apply a negative
(positive) field to the surface. 
The peak mainly comes from the quasiparticle having momentum of $\tk$$\simeq$$\pi/2$
as discussed before, which acquire an energy shift $-$$e\vfy A_y$.
In the case of  negative $B_{\rm ext}$,
there is an additional positive contribution to the  vector potential
which shifts the peak related to the gap towards higher energies.
If $B_{\rm ext}$ is reversed,
the vector potential decreases and
is almost zero at the surface when $B_{\rm ext}$=0.1$B_{\rm C}$.
Then the two peaks merge into a single peak as shown in Fig. \ref{fig:8}(d).
Increasing the external field further in the positive direction,
the vector potential is reversed (see Fig. \ref{fig:8}(c)),
so that the peak splits again.
The lines A and B in Fig. \ref{fig:8}(d) correspond to the peak
with the positive and negative $k_y$,
respectively.
Energetically the case of the negative field is expected to be stable
as it further opens the gap.

In the $d$-wave case bound states are located at $E$=0,
\cite{Hu}
so that the peak structure at the zero energy is split by the vector potential.
\cite{Fogelstrom}
In the present $p_x$$\pm$i$p_y$-wave case bound states distribute
inside the energy gap 
and form an almost flat local density of states at the surface as
shown in Fig. \ref{fig:4}(d). 
Near zero energy the shift is small due to the small $\vfy$ factor,
since the states near zero energy are formed by the quasiparticles having
momenta of $\tk$$\ll$$\pi/2 $. 
Therefore the energy shift due to an external field 
is not observable for energies below the gap.
The gap structure sensitive to the applied external magnetic field
is located rather at energies corresponding to the enhanced $\Delta_y$ at
the surface. 

\section{Domain Wall}
A modified quasiparticle spectrum is also expected near the domain wall.
Clearly, domain walls cost energy and are not desirable modification
of the order parameter. However, they exist even at low temperatures,
since once they have formed at the onset of superconductivity,
they are easily pinned at defects in the
material and cannot move out of the sample. 
Domain walls between the pinning centers may give good targets for
scanning tunneling spectroscopy on the sample surface. 
Physical properties of domain walls in unconventional superconductors
have been investigated in the framework of the GL theory.
\cite{Volovic,Sigrist4}
Here we investigate their quasiparticle properties within the
quasi-classical approximation which is essential for the tunneling
spectroscopy.  
We will not examine effects of the external magnetic field in this case,
because the magnetic field would not reach the domain wall 
due to the Meissner screening.

There are two kinds of domain walls separating the two degenerate
superconducting states. For the domain wall perpendicular to the $
x$-axis we have the ($p_x$$-$i$p_y$$\mid$$p_x$+i$p_y$) structure  and 
the ($-$$p_x$+i$p_y$$\mid$$p_x$+i$p_y$) structure, i.e. one of the two order
parameter components should change sign at the domain wall. 
From the GL formulation it is expected that the first type is more stable,
since the gradient free energy of the $p_x$-wave is three
times larger 
than that of $p_y$-wave (see $F_4^{\rm grad}$ in eq. (\ref{eqn:A.11})).

\begin{figure}[h]
\begin{center}
\input{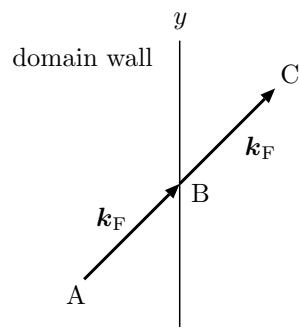}
\end{center}
\caption{
Classical trajectory of a quasiparticle going through a domain wall,
in which the incident momentum is conserved.
}
\label{fig:1b}
\end{figure}

We assume that there is no scattering at the domain wall leading to
quasi-classical trajectories as shown in Fig. \ref{fig:1b}.
The boundary condition for the domain wall case
is also given by eq. (\ref{eqn:2.11}) with ${\bkf}_1$=${\bkf}_2$.
In the domain wall case $J_y(x)$ is symmetric under the $x$$\rightarrow$$-$$x$ transformation.
Therefore the vector potential is also symmetric and
the magnetic field becomes anti-symmetric.

\subsection{The ($p_x$$-$i$p_y$$\mid$$p_x${\rm +i}$p_y$) domain wall}
\begin{figure}
\begin{center}
\begin{minipage}[t]{7cm}
\epsfxsize=7cm
\epsfbox{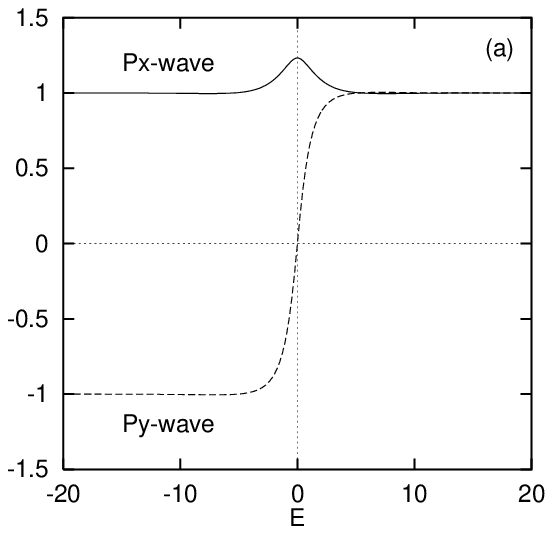}
\end{minipage}
\begin{minipage}[t]{7cm}
\epsfxsize=7cm
\epsfbox{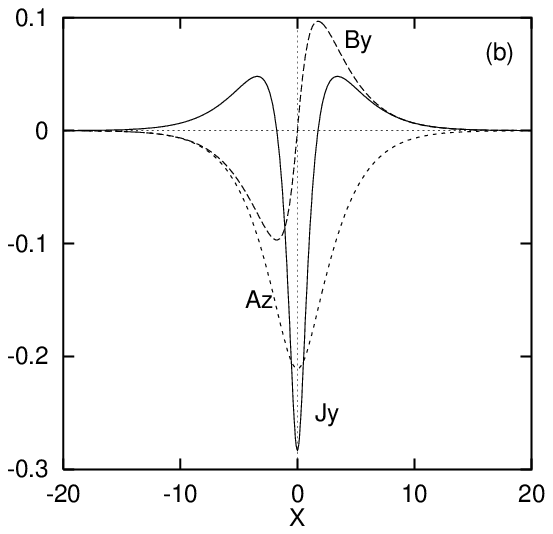}
\end{minipage}
\begin{minipage}[t]{9cm}
\epsfxsize=9cm
\epsfbox{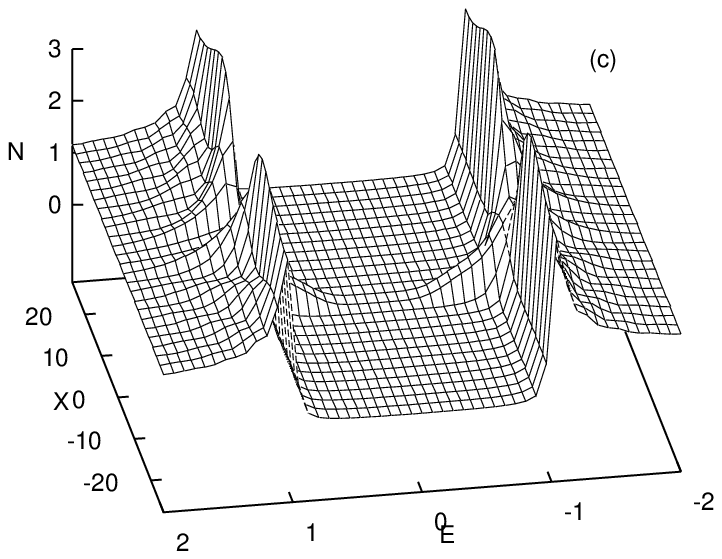}
\end{minipage}
\begin{minipage}[t]{7cm}
\epsfxsize=7cm
\epsfbox{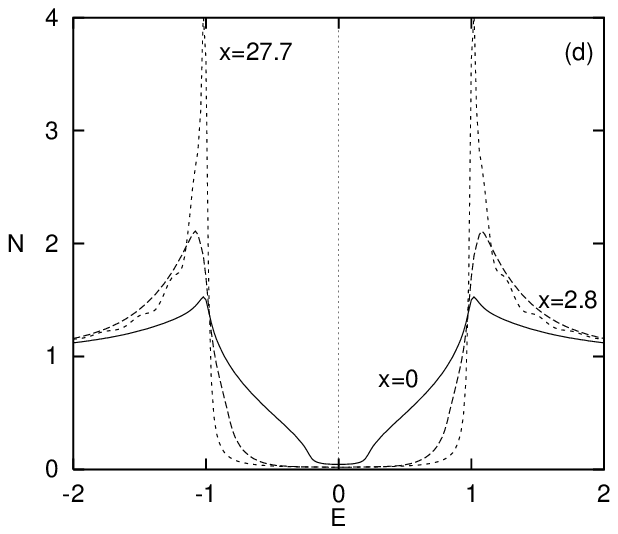}
\end{minipage}
\end{center}
\caption{
(a)~Spatial dependence of the self-consistent order parameters for $p_x$- and $p_y$-waves.
$\Delta_x$ and $\Delta_y$ are real and imaginary,
respectively.
The set of parameters are chosen as $T$=0.2$T_{\rm C}$, $\omega_{\rm C}$=10$T_{\rm C}$,
$\kappa$=2.5.
(b)~The self-consistent current density $J_y(x)$, the magnetic field $B_z(x)$
and the vector potential $A_y(x)$.
They are scaled by $J_0$=$e\vf N(0)T_{\rm C}$,
$B_{\rm C}$=$\Phi_0/2\sqrt{2}\pi\xi_0\lambda_{\rm L}$ and $\Delta(0)/e\vf$,
respectively.
(c)~Total local density of states.
(d)~Total local density of states at various positions.
They are symmetric under $x$$\rightarrow$$-$$x$.
}
\label{fig:5}
\end{figure}

In Fig. \ref{fig:5} we show the results of the self-consistent solution.
A current is induced near the domain wall and it is screened due to
the  Meissner effect.\cite{Sigrist4}
Figure \ref{fig:5}(a) shows that the relative phase between the $p_x$- and
$p_y$-wave component suddenly changes, if we go through the domain
wall (type I). On the other hand, the relative phase
can change continuously, approximately as ${\pi \over 2}{\rm tanh}(x/r\xi_0)$
(type II). These two types were discussed previously within the GL
formulation. \cite{Volovic,Sigrist4}
In the case of type II domain walls the relative phase changes
gradually from +$\pi/2$ to $-$$\pi /2$. Our calculation shows that
the type I solution is energetically favored for the cylindrical
symmetry Fermi surface case. 

\begin{figure}
\begin{center}
\input{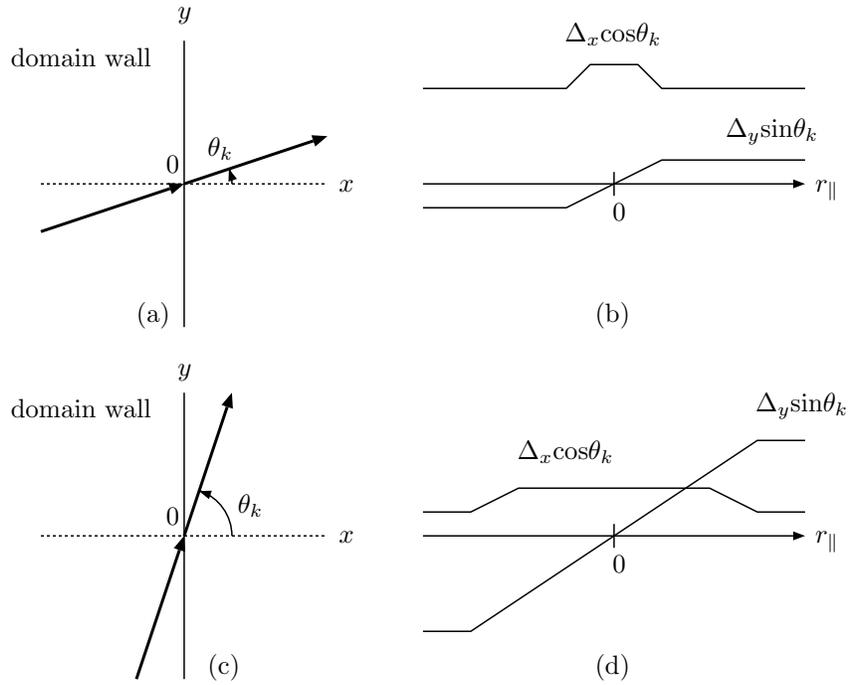}
\end{center}
\caption{
(a)~A classical trajectory of an quasiparticle going through a domain wall.
(b)~A schematic spatial dependence of the self-consistent order parameters
along the classical trajectory.
$\rpara$ is taken along the classical trajectory.
(c)~Same as (a) for a larger $\tk$.
(d)~Same as (b) for a larger $\tk$.
}
\label{fig:6}
\end{figure}

Despite several similarities with the order parameter structure
between the surface and the domain wall, there
are important differences in the quasiparticle spectrum as seen by
comparing Figs. \ref{fig:4}(d) and \ref{fig:5}(d).
In Fig. \ref{fig:4}(d) a gap structure can be seen at
$E/\Delta(T)$$\simeq$1.25, while there is no gap feature at the
corresponding energy in Fig.\ref{fig:5}(d).
In Fig. \ref{fig:6} we show the schematic spatial dependence
of the self-consistent order parameters,
which a quasiparticle encounters along the classical trajectory.
As we mentioned in $\S$3,
the higher energy peak for the surface case is mainly formed by the
quasiparticles 
which run almost parallel to the surface.
The same situation occurs also in the domain wall case,
where the quasiparticle with momentum $\tk$$\simeq$0 mainly contribute to
the gap structure. 
Compared to the surface cases the distance in which the quasiparticle experiences
the enhanced order parameter $\Delta_x$ is short (see Fig. \ref{fig:6}(b)),
so that the gap structure does not appear
in the ($p_x$$-$i$p_y$$\mid$$p_x$+i$p_y$) domain wall case.

There is another characteristic point in Fig. \ref{fig:5}(d), 
a small gap near the zero energy.
The density of states near the zero energy mainly comes from the region
where $\tk$$\simeq$$\pi/2$ for the domain wall case (see
Figs. \ref{fig:6}(c) and (d)).
The vector potential can generate a shift of the energy towards higher
values. In the $\tk$$\simeq$$\pi/2$ case
the magnitude of the shift can be estimated as $e\vf A_y(0)$$\simeq$0.1,
which is shown in Fig. \ref{fig:5}(b).
The small gap in Fig. \ref{fig:5}(d) is about 0.1,
which is consistent to this estimate.
In fact, neglecting the vector potential,
we find that the gap structure in the small energy region has disappeared.

A further difference from the surface case is the v-shaped density of states
below the bulk energy gap.
The origin of this feature is the gradual spatial dependence of
$\Delta_y\sin\tk$ on the classical trajectory 
with $\tk$$\simeq$$\pi/2$ (see Fig. \ref{fig:6}(d)),
which gives rise to additional bound states close to the bulk gap energy,
\cite{Matsumoto1}
so that the density of states at low energies is shifted up to higher energy region.

\subsection{The ($-$$p_x${\rm +i}$p_y$$\mid$$p_x${\rm +i}$p_y$) domain wall}
In this case the system also favors the type I behavior for the
relative phase. Here the situation is very similar to the surface case.
(In the surface case $\Delta_x$ changes its sign after the reflection
at surface.) Therefore most of the properties are the same and
differences between them appear only on the Fermi wave length 
scale. 

If the Fermi surface is not cylindrical, then the
properties of the domain walls can be modified
and a type II domain wall may be realized in some cases.
\cite{Sigrist4}

\section{Summary}

In this paper we analyzed the properties of the order parameter and
the quasiparticles of a $p_x$$\pm$i$p_y$-wave superconductor near the
surface and at a domain wall. For this purpose we used a
quasiclassical approximation which allows us to determine 
selfconsistently the current density, magnetic field and vector
potential distribution besides the order parameter. 

The spatial dependence of the order parameter at the surface shows a
suppression for the component $\bp\parallel\bn$
where $\bn$ is the surface normal vector, i.e. the $ p_x $-component is
suppressed for the surface perpendicular to $ x $-axis. The
perpendicular component, on the other hand, is slightly enhanced. 
In the surface region we found bound states whose energy strongly depends on 
the position of the quasiparticle momentum on the Fermi surface. The 
average over the Fermi surface shows that this bound state yields a
density of states which is essentially constant 
within the gap at the surface. As we move towards the bulk region
these bound states gradually disappear and the complete bulk gap is
recovered. We have also demonstrated 
that an external field (parallel to the c-axis) modifies the local
density of states differently for the two degenerate states. 
Similar to the case of the time reversal symmetry breaking surface
of the $d$-wave superconductor we find that the bound states carry a
spontaneous current whose direction is opposite for the two degenerate
superconducting bulk states. Furthermore an external magnetic field
modifies the local density of states at the surface significantly only 
for energies slightly above the bulk energy gap, while the subgap
state structure is essentially unaffected. 

The domain wall shows various similarities with the surface.
One order parameter component is suppressed while the other is slightly enhanced.
In contrast to the situation at surface, however, the
suppressed and enhanced components are
the ones with $\bp\perp\bn$ and $\bp\parallel\bn$, respectively,
which yields the narrower (energetically more favorable) domain wall than in the
opposite case. At the surface the choice is not available since the
role of the order parameter components is entirely determined by the
boundary condition due to the surface scattering.
There are, in principle, two types of domain walls: In the type I domain
wall the suppressed order parameter component vanishes in the center
of the domain wall. For the type II domain wall this component
introduces a phase twist which yields a finite modulus everywhere.
The latter domain wall state is two-fold degenerate. The calculation
using a cylindrical Fermi surface showed that the type I domain wall
is more stable at all temperatures. This would change, if, for
example, the anisotropy of the Fermi surface is taken into account. 

The local density of states in the center of the type I domain wall
shows a tiny gap, while the structure of the large bulk gap remains
visible. The shift of density of states is due to the presence of
bound states in the domain wall similar to the surface and the tiny
gap is a feedback effect of the vector potential created by the
quasiparticles themselves. This change of 
the density of states could provide a possible way to detect domain walls by 
scanning tunneling microscopy. For a given voltage within the bulk gap 
the scanning of the (c-axis oriented) sample surface would yield an
enhanced current in the region close the domain wall. 

Analogous to the surface case the bound states in the domain wall
carry a spontaneous current which yields a magnetic field
distribution. In contrast to the surface the net magnetization
generated by these currents at the domain wall has to be zero, since
inside the superconductor magnetic flux can only enter in form of
vortices enclosing a fixed quantum of flux. Thus, the magnetic fields
of the domain walls are invisible on a macroscopic level. Even 
a scanning SQUID microscope has, at present, still too little spatial
resolution to observe these fields. 

For the cylindrical Fermi surface the properties of the domain wall
and the surface does not depend on the orientation of the normal
vector as long as it lies within the $ x $-$y$-plane. The real
superconductor has certainly an anisotropic Fermi surface which could
change various properties of the quasiparticle states for both the
surface and the domain wall. To what degree changes occur remains a
problem for future studies.

\section*{Acknowledgments}
One of the authors (M. M.) expresses his sincere thanks to Prof. H. Shiba
for his critical reading of the manuscript.
We are greateful to Dr. S. Higashitani and Prof. K. Nagai
for pointing out the correct expression of the current density.
We would like to thank also Dr. Y. Okuno for many helpful discussions.
This work was supported by Grant-in-Aid for Scientific Research
form the Ministry of Education, Science and Culture, Japan.

\appendix
\section{Ginzburg-Landau Free Energy}
In $\S$3 we have shown that the self-consistent order parameters
favor the time-reversal symmetry breaking state ($p_x$$\pm$i$p_y$-wave),
and that the $p_y$-wave part is enhanced by the presence of the $p_x$-wave.
In order to understand these results from a different point of view,
we derive the GL free energy.
The GL free energy can be easily derived by using the quasi-classical
Green function.\cite{Schopohl2}
For simplicity the vector potential is neglected here.

To derive the GL free energy,
let us rewrite eq. (\ref{eqn:2.6}) in the following form:
\beqn
\om f(\bkf,\om,x)&=&\Delta(\bkf,x)g(\bkf,\om,x)-Df(\bkf,\om,x), \cr
\om \fbar(\bkf,\om,x)&=&\Delta^*(\bkf,x)g(\bkf,\om,x)+D\fbar(\bkf,\om,x), \cr
g^2(\bkf,\om,x)&=&1-f(\bkf,\om,x)\fbar(\bkf,\om,x),
\label{eqn:A.1}
\eeqn
where $D$=${\vfx \over 2}{\del \over \del x}$
and we have assumed the system is uniform in the $y$ direction,
i.e. there is no $y$ dependence in $g$, $f$ and $\fbar$.
$\Delta(\bkf,x)$ is assumed to be composed of $p_x$- and $p_y$-waves.
Since we are interested in the temperature near $T_{\rm C}$,
the order parameters are assumed to be small.
The gradient term ($D$ term) can be expressed as
\beq
{D \over T}\sim{\vert\vfx\vert \over T_{\rm C}}{\del \over \del x}
\sim{\vert\vfx\vert \over \Delta(0)}{\del \over \del x}
\sim{\del \over \del(x/\xi_0)},
\eeq
where $\xi_0$=$\vf/\pi\Delta(0)$, which is the coherence length at $T$=0.
For $T$$\rightarrow$$T_{\rm C}$,
the spatial dependence of the order parameter is of the order of
$\xi(T)$$\sim$$\vf/\pi\Delta(T)$, and  $D/T$$\sim$$O(\Delta/T)$$\ll$1.
Let us treat $\Delta/T$ as a perturbation and derive the functions
$f$, $\fbar$ and $g$ 
in a power series of $\Delta/T$, 
\beqn
f&=&f_0+f_1+f_2+\ldots, \cr
\fbar&=&\fbar_0+\fbar_1+\fbar_2+\ldots, \cr
g&=&g_0+g_1+g_2+\ldots,
\eeqn
where the subscripts represent the order of $\Delta/T$.
The lowest order is contained in the bulk solution
\beq
f_{\rm bulk}={\Delta \over \sqrt{\om^2+\vert\Delta\vert^2}},~~~
\fbar_{\rm bulk}={\Delta^* \over \sqrt{\om^2+\vert\Delta\vert^2}},~~~
f_{\rm bulk}={\om \over \sqrt{\om^2+\vert\Delta\vert^2}}.
\eeq
Therefore the zeroth order functions are given by
\beq
f_0=0,~~~\fbar_0=0,~~~g_0={\om \over \vert\om\vert}.
\label{eqn:A.5}
\eeq
Since $f$ and $\fbar$ are related to the order parameter,
they do not have the zeroth order term.
We can obtain the functions $f$, $\fbar$ and $g$ up to any order of $\Delta/T$
by using eqs. (\ref{eqn:A.1}) and (\ref{eqn:A.5}).
\cite{Schopohl2}
The results are as follows:
\begin{eqnarray}
&&\left\{
\begin{array}{ll}
f_1 = {\Delta \over \vert\om\vert}, \cr
\fbar_1 ={\Delta^* \over\vert\om\vert}, \cr
g_1 = 0,
\end{array}
\right. \cr
&&\left\{
\begin{array}{ll}
f_2 =-{D\Delta \over \om\vert\om\vert}, \cr
\fbar_2 = {D\Delta^* \over \om\vert\om\vert}, \cr
g_2 =-{\vert\Delta\vert^2 \over 2\om\vert\om\vert},
\end{array}
\right. \cr
&&\left\{
\begin{array}{ll}
f_3 = {-\Delta\vert\Delta\vert^2/2+D^2\Delta \over \om^2\vert\om\vert}, \cr
\fbar_3 = {-\Delta^*\vert\Delta\vert^2/2+D^2\Delta^* \over \om^2\vert\om\vert}, \cr
g_3 = {-\Delta D\Delta^*+\Delta^* D\Delta \over 2\om^2\vert\om\vert}.
\end{array}
\right. \\
&&\left\{
\begin{array}{ll}
f_4 = {(-\Delta^2 D\Delta^*+\vert\Delta\vert^2D\Delta)/2+D\Delta\vert\Delta\vert^2/2-D^3\Delta \over \om^3\vert\om\vert}, \cr
\fbar_4 = {({\Delta^*}^2 D\Delta-\vert\Delta\vert^2D\Delta^*)/2-D\Delta^*\vert\Delta\vert^2/2+D^3\Delta^* \over \om^3\vert\om\vert}, \cr
g_4 = {3\vert\Delta\vert^4/4-\Delta D^2\Delta^*-\Delta^* D^2\Delta+(D\Delta)(D\Delta^*)\over  2\om^3\vert\om\vert},
\end{array}
\right. \cr
&&\left\{
\begin{array}{ll}
f_5 = {
      \Delta\bigl[
                   3\vert\Delta\vert^4/4-\Delta D^2\Delta^*-\Delta^* D^2\Delta+(D\Delta)(D\Delta^*)
            \bigr]
          -D\bigl[
              -\Delta^2 D\Delta^*+\vert\Delta\vert^2 D\Delta+D\Delta\vert\Delta\vert^2
              -2D^3\Delta
            \bigr]
        \over 2\om^4\vert\om\vert}, \cr
\fbar_5 = {
     \Delta^*\bigl[
                   3\vert\Delta\vert^4/4-\Delta^* D^2\Delta-\Delta D^2\Delta^*+(D\Delta^*)(D\Delta)
             \bigr]
     -D\bigl[
              -{\Delta^*}^2 D\Delta+\vert\Delta\vert^2 D\Delta^*+D\Delta^*\vert\Delta\vert^2
              -2D^3\Delta^*
       \bigr]
         \over 2\om^4\vert\om\vert}, \cr
\end{array}
\right.
\nonumber
\end{eqnarray}
The gap equation can be written by using $f$ and $\fbar$ as
\beq
\left(
  \matrix{
    \Delta_x(x) \cr
    \Delta_y(x) \cr
  }
\right)
= 2\pi V_p T\sum_{0<\om<\omega_{\rm C}}{1 \over 2\pi} \int_{-\pi}^\pi {\rm d}\tk
\left(
  \matrix{
    2\cos(\tk) \cr
    2\sin(\tk) \cr
  }
\right)
\bigl[ f(\tk,\om,x)+\fbar^*(\tk\,\om,x) \bigr].
\label{eqn:3.70}
\eeq
Substituting $f$ and $\fbar$ up to the fifth order into eq. (\ref{eqn:3.70}),
we obtain the gap equation for the $p_x$-wave
\begin{eqnarray}
&&-A_2\Delta_x
+A_4
  \bigl[
    3\Delta_x\vert\Delta_x\vert^2+2\Delta_x\vert\Delta_y\vert^2+\Delta_x^*\Delta_y^2
  \bigr]
- 6A_4 \dbar^2\Delta_x \cr
&&- {3 \over 4}A_6
\bigl[
      5\Delta_x\vert\Delta_x\vert^4
     +\Delta_x^3{\Delta_y^*}^2
     +6\Delta_x\vert\Delta_x\vert^2\vert\Delta_y\vert^2
     +3\Delta_x^*\vert\Delta_x\vert^2\Delta_y^2
     +3\Delta_x\vert\Delta_y\vert^4
     +2\Delta_x^*\vert\Delta_y\vert^2\Delta_y^2
\bigr] \cr
&&+ 5A_6
   \bigl[
    \Delta_x^2\dbar^2\Delta_x^*
    +2\Delta_x\vert\dbar\Delta_x\vert^2
    +3\Delta_x^*(\dbar\Delta_x)^2
    +4\vert\Delta_x\vert^2\dbar^2\Delta_x
  \bigr]
- 10A_6 \dbar^4\Delta_x \cr
&&+ A_6
   \bigl[
    2\Delta_x\Delta_y\dbar^2\Delta_y^*
   +4\Delta_x\Delta_y^*\dbar^2\Delta_y
   +2\Delta_x\vert\dbar\Delta_y\vert^2
   +4\Delta_x^*\Delta_y\dbar^2\Delta_y
   +3\Delta_x^*(\dbar\Delta_y)^2
   +\Delta_y^2\dbar^2\Delta_x^* \cr
&&~~~~~~
+4\vert\Delta_y\vert^2\dbar^2\Delta_x
   +2\Delta_y(\dbar\Delta_x)(\dbar\Delta_y^*)
   +2\Delta_y(\dbar\Delta_y)(\dbar\Delta_x^*)
   +6\Delta_y^*(\dbar\Delta_x)(\dbar\Delta_y)
  \bigr]=0, \cr
&&A_2=1-{T \over T_{\rm C}},~~~
A_4={1 \over 4}{7 \over 8}{\zeta(3) \over (\pi T)^2},~~~
A_6={1 \over 8}{31 \over 32}{\zeta(5) \over (\pi T)^4},~~~
\dbar={1 \over 2}\vf{\partial \over \partial x},
\end{eqnarray}
and for the $p_y$-wave
\begin{eqnarray}
&&-A_2\Delta_y
+A_4
  \bigl[
    3\Delta_y\vert\Delta_y\vert^2+2\Delta_y\vert\Delta_x\vert^2+\Delta_y^*\Delta_x^2
  \bigr]
- 2A_4 \dbar^2\Delta_y \cr
&&- {3 \over 4}A_6
\bigl[
      5\Delta_y\vert\Delta_y\vert^4
     +\Delta_y^3{\Delta_x^*}^2
     +6\Delta_y\vert\Delta_y\vert^2\vert\Delta_x\vert^2
     +3\Delta_y^*\vert\Delta_y\vert^2\Delta_x^2
     +3\Delta_y\vert\Delta_x\vert^4
     +2\Delta_y^*\vert\Delta_x\vert^2\Delta_x^2
\bigr] \cr
&&+ A_6
   \bigl[
    \Delta_y^2\dbar^2\Delta_y^*
    +2\Delta_y\vert\dbar\Delta_y\vert^2
    +3\Delta_y^*(\dbar\Delta_y)^2
    +4\vert\Delta_y\vert^2\dbar^2\Delta_y
  \bigr]
- 2A_6 \dbar^4\Delta_y \cr
&&+ A_6
   \bigl[
    2\Delta_y\Delta_x\dbar^2\Delta_x^*
   +4\Delta_y\Delta_x^*\dbar^2\Delta_x
   +2\Delta_y\vert\dbar\Delta_x\vert^2
   +4\Delta_y^*\Delta_x\dbar^2\Delta_x
   +3\Delta_y^*(\dbar\Delta_x)^2
   +\Delta_x^2\dbar^2\Delta_y^* \cr
&&~~~~~~
+4\vert\Delta_x\vert^2\dbar^2\Delta_y
   +2\Delta_x(\dbar\Delta_y)(\dbar\Delta_x^*)
   +2\Delta_x(\dbar\Delta_x)(\dbar\Delta_y^*)
   +6\Delta_x^*(\dbar\Delta_y)(\dbar\Delta_x)
  \bigr]=0. \nonumber \\
\end{eqnarray}
Here $\zeta(z)$ is the Riemann zeta function.
The GL free energy should recover the above gap equations
through the following Euler-Lagrange equation,
\begin{equation}
\Bigl[
{\partial \over \partial \Delta^*}
  - \dbar{\partial \over \partial (\dbar\Delta^*)} 
  + \dbar^2{\partial \over \partial (\dbar^2{\Delta^*}^2)}
\Bigr] F =0.
\end{equation}
The result of the free energy density up to sixth order is as follows:
\begin{eqnarray}
F &\propto& F_2+F_4+F_6+F_4^{\rm grad}
      +F_{6{\rm a}}^{\rm grad}+F_{6{\rm b}}^{\rm grad}+F_{6{\rm c}}^{\rm grad}, \cr
F_2 &=& -A_2\bigr[\vert\Delta_x\vert^2+\vert\Delta_y\vert^2\bigr], \cr
F_4 &=& A_4
  \bigl[
    {3 \over 2}(\vert\Delta_x\vert^4+\vert\Delta_y\vert^4)
    +2\vert\Delta_x\vert^2\vert\Delta_y\vert^2
    +{1 \over 2}(\Delta_x^2{\Delta_y^*}^2+{\Delta_x^*}^2\Delta_y^2)
  \bigr], \cr 
F_6 &=& -A_6
  \bigl[
    {5 \over 4}(\vert\Delta_x\vert^6+\vert\Delta_y\vert^6)
    +{9 \over 4}(\vert\Delta_x\vert^4\vert\Delta_y\vert^2
                +\vert\Delta_x\vert^2\vert\Delta_y\vert^4)
    +{3 \over 4}(\vert\Delta_x\vert^2+\vert\Delta_y\vert^2)
                (\Delta_x^2{\Delta_y^*}^2+{\Delta_x^*}^2\Delta_y^2)
  \bigr], \cr
F_4^{\rm grad} &=& A_4
  \bigl[
    6\vert\dbar\Delta_x\vert^2+2\vert\dbar\Delta_y\vert^2
  \bigr], \cr
F_{6{\rm a}}^{\rm grad} &=& -A_6
  \Bigl\{
    20\vert\Delta_x\vert^2\vert\dbar\Delta_x\vert^2
   +{5 \over 2}
    \bigl[
      \Delta_x^2(\dbar\Delta_x^*)^2+{\Delta_x^*}^2(\dbar\Delta_x)^2
    \bigr] \cr
&&~~~~~~~~   +4\vert\Delta_y\vert^2\vert\dbar\Delta_y\vert^2
   +{1 \over 2}
    \bigl[
      \Delta_y^2(\dbar\Delta_y^*)^2+{\Delta_y^*}^2(\dbar\Delta_y)^2
    \bigr]
  \Bigr\}, \cr
F_{6{\rm b}}^{\rm grad} &=& -A_6
  \bigl[
    10\vert\dbar^2\Delta_x\vert^2+2\vert\dbar\Delta_y\vert^2
  \bigr], \cr
F_{6{\rm c}}^{\rm grad} &=& -A_6
  \Bigl\{
    {1 \over 2}
      \bigl[
        \Delta_x^2(\dbar\Delta_y^*)^2+{\Delta_x^*}^2(\dbar\Delta_y)^2
       +\Delta_y^2(\dbar\Delta_x^*)^2+{\Delta_y^*}^2(\dbar\Delta_x)^2
      \bigr] \cr
&&~~~~~~
   +2 \bigl[
        \Delta_x\Delta_y(\dbar\Delta_x^*)(\dbar\Delta_y^*)
       +\Delta_x^*\Delta_y^*(\dbar\Delta_x)(\dbar\Delta_y)
      \bigr] \cr
&&~~~~~~
   +4 \bigl[
        \Delta_x\Delta_y^*(\dbar\Delta_x)(\dbar\Delta_y^*)
       +\Delta_x^*\Delta_y(\dbar\Delta_x^*)(\dbar\Delta_y) \cr
&&~~~~~~~~~
       +\Delta_x\Delta_y^*(\dbar\Delta_x^*)(\dbar\Delta_y)
       +\Delta_x^*\Delta_y(\dbar\Delta_x)(\dbar\Delta_y^*)
       +\vert\Delta_x\vert^2\vert\dbar\Delta_y\vert^2
       +\vert\Delta_y\vert^2\vert\dbar\Delta_x\vert^2
      \bigr]
  \Bigr\}.
\nonumber \\
\label{eqn:A.11}
\end{eqnarray}
The coefficient of the free energy density is chosen
to reproduce the well known bulk GL free energy.
The terms which suppress the spatial change of the order parameters are
$6A_4\vert\dbar\Delta_x\vert^2$ and $2A_4\vert\dbar\Delta_y\vert^2$
in the fourth gradient term.
Therefore, the spatial variation of $ \Delta_x $ is energetically more 
expensive than that of $ \Delta_y $ along the $x$-direction. 
The asymmetry of the $p_x$- and $p_y$-waves comes from the $D$ term,
since $D$$\propto$$\vfx$=$\vf\cos\tk$.
$F_{6{\rm a}}^{\rm grad}$ and $F_{6{\rm b}}^{\rm grad}$ are the sixth order gradient free energies
in which $\Delta_x$ does not couple with $\Delta_y$,
while $F_{6{\rm c}}^{\rm grad}$ is the free energy in which the two
components are coupled.


\clearpage

\begin{figure}[h]
\begin{center}
\input{figure1.tps}
\end{center}
\caption{
(a)~Classical trajectory of a quasiparticle,
in which the momentum of the incident and reflected quasiparticles
along the surface is conserved.
${\bkf}_1$ and ${\bkf}_2$ are the momentum with $\kfx$$<$0 and $\kfx$$>$0,
respectively.
(b)~$\rpara$ is taken along the classical trajectory.
$\Delta({\bkf}_1,\rpara)$ and $\Delta({\bkf}_2,\rpara)$ are the order parameters
for the incident and reflected quasiparticles,
respectively.
A schematic spatial dependence is plotted.
}
\label{fig:1}
\end{figure}

\end{document}